\documentclass[aps,prl,showpacs,groupedaddress,twocolumn,preprintnumbers,amsmath,amssymb]{revtex4}

\usepackage{graphicx}
\usepackage{dcolumn}
\usepackage{bm}
\usepackage{braket}

\begin{document}

\title{Projective measurement of a single nuclear spin qubit by using two-mode cavity QED}

\author{Yujiro Eto$^{1,\dag}$}
\author{Atsushi Noguchi$^{1,2}$}
\author{Peng Zhang$^{1,\ddag}$}
\author{Masahito Ueda$^{1,3}$}
\author{Mikio Kozuma$^{1,2}$}
\affiliation{%
$^{1}$ERATO Macroscopic Quantum Control Project, JST, 2-11-16 Yayoi, Bunkyo-Ku, Tokyo 113-8656, Japan}
\affiliation{%
$^{2}$Department of Physics, Tokyo Institute of Technology, 2-12-1 O-okayama, Meguro-ku, Tokyo 152-8550, Japan}
\affiliation{%
$^{3}$Department of Physics, University of Tokyo, Hongo, Bunkyo-ku, Tokyo 113-0033, Japan}
\altaffiliation[Present affiliation:]{$\dag$ National Institute of Information and Communications Technology, $\ddag$ Renmin University of China}

\date{\today}
             
\begin{abstract}
We report the implementation of projective measurement on a single 1/2 nuclear spin of the $^{171}$Yb atom
by measuring the polarization of cavity-enhanced fluorescence.
To obtain cavity-enhanced fluorescence having a nuclear-spin-dependent polarization,
we construct a two-mode cavity QED system,
in which two cyclic transitions are independently coupled to each of the orthogonally polarized cavity modes,
by manipulating the energy level of $^{171}$Yb.
This system can associate the nuclear spin degrees of freedom with the polarization of photons, which will facilitate the development of hybrid quantum systems.

\end{abstract}

\pacs{03.67.Lx, 42.50.Ex, 42.50.Pq}
\maketitle
Alkaline-earth-like atoms are considered as a promising candidate for use as a robust qubit in quantum information science and for the realization of atomic clocks \cite{Ye08} 
owing to their unique energy-level structures.
For fermionic species, the nuclear spins in the $^{1}S_{0}$ ground state can store quantum information while avoiding the decoherence caused by the magnetic field fluctuation.
The presence of $^{3}P$ metastable excited states opens up further possibilities for the optical manipulation of quantum coherence \cite{Reichenbach07,Reichenbach09}.
By harnessing these features,
novel schemes for quantum computing \cite{Daley08,Gorshkov09,Shibata09} and a quantum simulator \cite{Gorshkov10} using an optical lattice have been proposed and extensively studied \cite{Fukuhara07}.

Another interesting approach is to combine a single atom with a cavity QED system, which provides an excellent control of the interactions between single atoms and single photons \cite{Kimble98,Vahala03}.
By using such a combination, nuclear-spin-dependent coupling with the light field can be realized,
thus enabling nuclear spin engineering using photons.
So far, by tuning the cavity resonance to the transition between one nuclear spin state of $^{1}S_{0}$ and one magnetic substate of $^{3}P_{1}$ in $^{171}$Yb,
fast detection with high fidelity \cite{Takeuchi10,Noguchi10} and cavity-enhanced Faraday rotation \cite{Takei10} have been demonstrated for single 1/2 nuclear spin.

In this Letter, we construct a two-mode cavity QED system with $^{171}$Yb, 
in which the two independent cyclic transitions in $^{1}S_{0} (I = 1/2)$-$^{3}P_{1} (F' = 3/2)$ are coupled to each of the orthogonally polarized cavity modes.
To avoid unwanted spin flips in the ground state, 
the energy levels of the magnetic substates of $^{3}P_{1} (F' = 3/2)$ are manipulated by irradiation with a beam that is slightly detuned from the $^{3}P_{1} (F' = 3/2)$-$^{3}D_{1} (F' = 1/2)$ transition.
By using this setup, the projective measurement on a single 1/2 nuclear spin is performed by detecting the polarization of the cavity-enhanced fluorescence.
In principle, our method enables us to not only distinguish between the spin up and down but also detect the absence of atoms in real time by monitoring the photon counts.
This setup can be applied to an entanglement beam splitter \cite{Hu09} to create the photon-spin, photon-photon, and spin-spin entanglements.
It can also be applied to achieve a nuclear-spin-dependent strong transition in the electron shelving technique \cite{Dehmelt82} used to determine clock transitions \cite{Wang07,Kohno09,Lemke09}.

\begin{figure}[b]
\includegraphics[width=7.5cm]{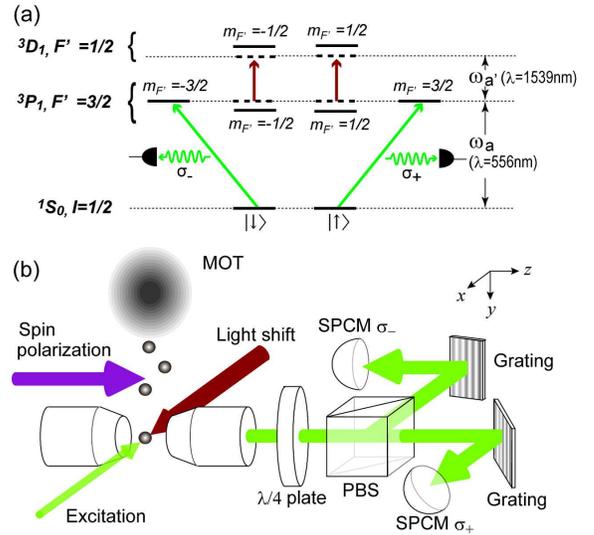}
\caption{(color online) 
(a) Relevant energy-level diagram of $^{171}$Yb. 
$\ket{\uparrow}$ and $\ket{\downarrow}$ denote the magnetic substates $m_I = \pm1/2$ in the ground state of $^{1}S_{0} (I = 1/2)$.
(b) Schematic of the experimental apparatus.
MOT: magneto-optical trap, $\lambda/4$ plate: quarter-wave plate, PBS: polarizing beam splitter, SPCM: single-photon-counting module.}
\label{fig1}
\end{figure}
 
Figure 1(a) shows the relevant energy levels of $^{171}$Yb.  
The magnetic substates $m_I = \pm1/2$ of the ground state $^{1}S_{0} (I = 1/2)$ are used as the nuclear spin qubit and are denoted as $\ket{\uparrow}$ and $\ket{\downarrow}$.
To avoid unwanted spin flips, 
the energy levels of $^{3}P_{1} (F' = 3/2)$ are manipulated using the ac Stark effect.
The substates  $m_{F'} = \pm1/2$ are shifted by irradiation with a $\pi$-polarized beam, 
which is nearly resonant with the $^{3}P_{1} (F' = 3/2)$-$^{3}D_{1} (F' = 1/2)$ transition (wavelength of 1539 nm).
The two substates $m_{I} = \pm1/2$ of $^{1}S_{0} (I = 1/2)$ are independently coupled to each of the cavity fields with $\sigma_{+}$ and  $\sigma_{-}$ polarizations. 
Therefore, a frequency-degenerate but polarization-nondegenerate two-mode cavity QED system is constructed.
Because the polarization of the cavity-enhanced fluorescence depends on the nuclear spin state, 
the projective measurements of $\ket{\uparrow}$ and $\ket{\downarrow}$ are accomplished by detecting the polarization of the emitted photons.

In our system, the transition between excited states $^{3}P_{1} (F' = 3/2)$-$^{3}D_{1} (F' = 1/2)$ is employed to manipulate the energy levels of $^{3}P_{1} (F' = 3/2)$ for the following reasons.
First, one can easily prevent the dark counts induced by the stray light from the light-shift beam, 
because the wavelength of the light-shift beam is different from that of the cavity-enhanced fluorescence.
Second, the nuclear spin states are not disturbed by the light-shift beam.

Our experimental setup is shown in Fig. 1(b).
$^{171}$Yb atoms are prepared by using Zeeman slowing and the double magneto-optical trapping techniques; 
for details, see Ref. \cite{Takeuchi10}.
After the atoms are released from the magneto-optical trap (MOT) located 7 mm above the microcavity,
the atoms are introduced into the cavity mode by gravity in 30 ms.
The intracavity atom number is adjusted by changing the loading time of the MOT.
During the free fall, each nuclear spin is prepared in $\ket{\uparrow}$ or $\ket{\downarrow}$ by the optical pumping with the $\sigma_{+}$ or $\sigma_{-}$ polarized beam, 
which is resonant with the $^{1}S_{0} (I = 1/2)$-$^{1}P_{1} (F' = 1/2)$ transition (wavelength of 399 nm).
Our cavity consists of two concave mirrors with a radius of curvature of 50 mm, and the TEM$_{00}$ waist size is 19 $\mu$m. 
The average transit time in the cavity is approximately 100 $\mu$s. 
The cavity length is stabilized to 150 $\mu$m by using FM sideband methods with a lock beam at 560 nm,
which is red detuned by four free spectrum ranges of the cavity from the resonance of the $^{1}S_{0} (I = 1/2)$-$^{3}P_{1} (F' = 3/2)$ transition (wavelength of 556 nm).
The relevant atom-cavity parameters are $(g_0, \kappa, \gamma)/2\pi = (2.8, 4.8, 0.091)$ MHz,
where $g_0$ is the maximum atom-cavity coupling rate,
$\kappa$ is the cavity decay rate [the half width at half maximum (HWHM) of the cavity resonance],
and $\gamma$ is the atom decay rate of the $^{1}S_{0} (I = 1/2)$-$^{3}P_{1} (F' = 3/2)$ transition (half of the natural line width).

The y-polarized excitation beam at 556 nm, which can be decomposed into $\sigma_{+}$ and $\sigma_{-}$ components, and the $\pi$-polarized light-shift beam at 1539 nm are shined simultaneously on the atoms in the microcavity in order to perform the projective measurement.
Here, the quantization axis is chosen to be in the z-direction.
The $\ket{\uparrow}$ or $\ket{\downarrow}$ state is excited, and photons are emitted into the cavity mode.
The emitted photons are detected by single-photon-counting modules [labeled as SPCM$\sigma_{+}$ and SPCM$\sigma_{-}$ in Fig. 1(b)],
after selecting the polarization of photons by using a quarter-wave plate ($\lambda/4$) and a polarizing beam splitter (PBS) 
and removing the lock beam by using the gratings.
All the photons emitted into the cavity mode are detected, with a total detection efficiency of $\eta_{t}=20$ $\%$ per scattered photon.
The dark counts at SPCM$\sigma_{+}$ and SPCM$\sigma_{-}$ are 1 and 0.5 ms$^{-1}$, respectively, which are mainly caused by the stray light from the lock beam. 

\begin{figure}[t]
\includegraphics[width=7.5cm]{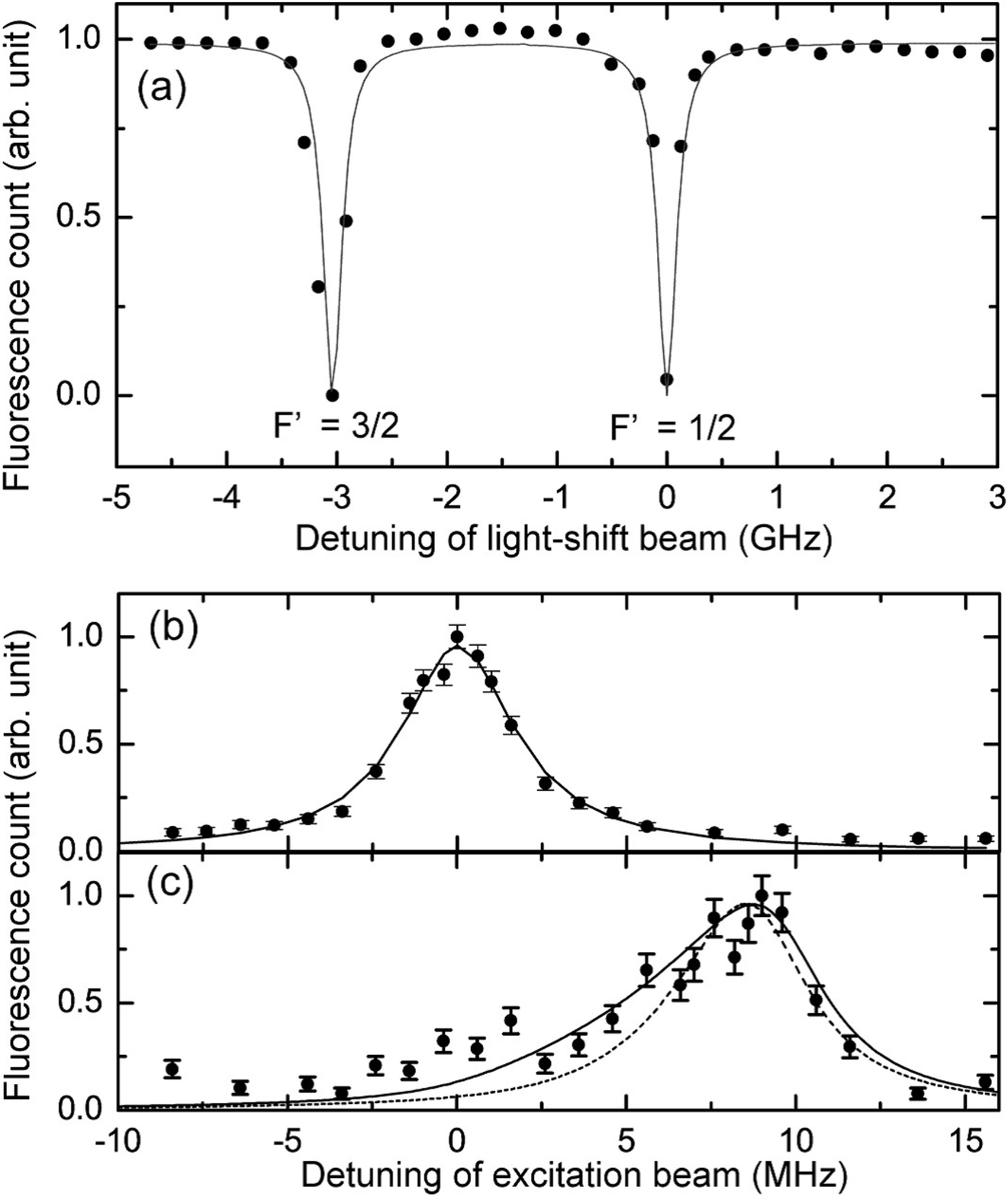}
\caption{(a) Hyperfine spectroscopy on the $^{3}D_{1}$ state of $^{171}$Yb. 
The fluorescence from the MOT are collected by a photomultiplier while shining the light-shift beam.
(b), (c) The cavity-enhanced fluorescence spectrum. 
The atoms are irradiated  (b) without and (c) with the light-shift beam.
The frequencies of both the excitation and the lock beam are simultaneously changed. 
}
\label{fig2}
\end{figure}

We carried out spectroscopy to tune the frequency of the light-shift beam to a value around the resonant frequency with the $^{3}P_{1} (F' = 3/2)$-$^{3}D_{1} (F' = 1/2)$ transition.
In this spectroscopy, the fluorescence at 556 nm scattered from the MOT is detected by a photomultiplier, and the beam at a  wavelength of 1539 nm is directed at the MOT.
Figure 2(a) shows the 556-nm fluorescence counts as a function of the detuning of the beam at a wavelength of 1539 nm,
whose intensity is 100 $\mu$W and the beam waist is a few millimeters.
As shown in Fig. 2(a), the photon counts are drastically reduced when the frequency is resonant on the hyperfine transition.
When the frequency is resonant on the hyperfine transition,
the trapped atoms are excited to the $^{3}D_{1}$ state and they decay to the $^{3}P_{0}$ state with a branching ratio of 64 \% \cite{DeMile95}.
Because the atoms in the $^{3}P_{0}$ state cannot be trapped by the MOT, the intensity of the fluorescence decreases dramatically.
The hyperfine resonances of the $^{3}P_{1} (F' = 3/2)$-$^{3}D_{1} (F' = 1/2)$ and $(F' = 3/2)$ transitions are identified from this measurement.

Though the decay rate of the $^{3}P_{1}$-$^{3}D_{1}$ transition is 16 kHz \cite{Loftus01}, wide dips are observed, as shown in Fig. 2(a) (the HWHM is approximately 100 MHz).
It should be noted that the power broadening caused by the probe beam is only of the order of a few megahertz.
This effect can be interpreted using a simple rate equation for the MOT atom number $N$, which is derived from the MOT loading rate $R$, the one-body decay rate $\Gamma_{0}$ (mainly due to the background gas collisions), and the scattering rate from the $^{3}D_{1}$ state $\Gamma_{1}$.
The rate equation is given by $\mathrm{d}N/\mathrm{d}t=R-(\Gamma_{0}+\eta\Gamma_{1})N$,
where $\eta$ is determined by multiplying the population of the $^{3}P_{1}$ state with the branching ratio to the $^{3}P_{0}$ state.
The steady-state solution thus becomes 
$N = R/(\Gamma_{0}+\eta\Gamma_{1})$.
In this equation, when $\eta\Gamma_{1}$ is comparable with $\Gamma_{0}$ ($\Gamma_{0}\sim0.5$ $s^{-1}$ in our experiment),
$N$ decreases dramatically.
From the steady-state solution,
the dip width is calculated to be $100$ MHz for a power density of 3 mW/cm$^2$; this power density is in reasonable agreement with our experimental value.  

Figure 2(b) and (c) shows the cavity-enhanced fluorescence spectra without and with irradiation, respectively.
Here, the power of the excitation beam is 1.8 $\mu$W.
The light-shift beam is set to the maximum available power of 9 mW, and the frequency is red detuned by 300 MHz from the $^{3}P_{1} (F' = 3/2)$-$^{3}D_{1} (F' = 1/2)$ transition. 
The beam waist size of the light-shift beam ($w_{l} \approx50$ $\mu$m) is set larger than that of the excitation beam ($w_{e} \approx25$ $\mu$m),
so that the light-shift beam reliably overlaps the excitation beam inside the cavity.
It should be noted that the transition strengths between $^{1}S_{0} (I = 1/2)$ and $m_{F'} = \pm3/2$ of $^{3}P_{1} (F' = 3/2)$ are three times larger than that between $^{1}S_{0}$ and $m_{F'} = \pm1/2$.
The observed fluorescence is mainly emitted from the $m_{F'} = \pm3/2$ states.
Thus, the shift in the spectrum in Fig. 2(c) reflects the shift in the energy level in $m_{F'} = \pm3/2$ of $^{3}P_{1} (F' = 3/2)$ and is mainly caused by the coupling between  $^{3}P_{1} (F' = 3/2)$ and $^{3}D_{1} (F' = 3/2)$.
A peak shift $\delta_{\pm 3/2}$ of $+8.5 \pm 1.0$ MHz is obtained from the spectrum in Fig. 2(c), 
which is reasonably consistent with the theoretically calculated value of $6.8$ MHz.
The light shift $\delta_{\pm 1/2}$ in $m_{F'} = \pm1/2$ of $^{3}P_{1} (F' = 3/2)$ is calculated to be $-16 \pm 1$ MHz by using the relationship between the transition strengths of hyperfine substates. 
From the values of $\delta_{\pm 1/2}$ and $\delta_{\pm 3/2}$, 
the frequency difference between $m_{F'} = \pm1/2$ and $m_{F'} = \pm3/2$ in $^{3}P_{1}$ is evaluated to be $24 \pm 2$ MHz.

While the observed spectrum in Fig. 2(b) shows a symmetrical distribution, 
a long tail appears on the left side of the distribution in Fig. 2(c).
We calculated the spectrum shape by considering the Gaussian intensity distribution of the light-shift beam, which induces the spatially varying light-shift strengths.
The dotted curve indicates the calculated spectrum shape for $w_{l} = 50$ $\mu$m used in this experiment, 
and it is quite different from the experimental results.
However, we can simulate the long tail of the distribution for the reduced $w_{l}$; that is, the solid curve is for  $w_{l} = 20$ $\mu$m.
We infer that the additional spatial intensity modulation is caused by the Fresnel diffraction by the concave mirrors or 
the imperfect overlap of the excitation beam with the light-shift beam due to the misalignment of the propagation axes.

\begin{figure}[t]
\includegraphics[width=7.5cm]{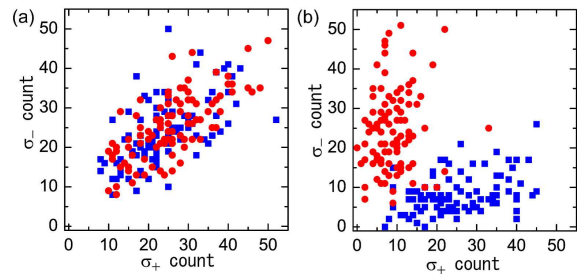}
\caption{(color online) Scattering diagrams of $\sigma_+$ and $\sigma_-$ counts (a) without and (b) with irradiation of the light-shift beam.
The blue squares (red circles) correspond to $\ket{\uparrow}$ ($\ket{\downarrow}$).
}
\label{fig3}
\end{figure}
  
To show that the unwanted spin flip rate in the ground state is reduced by the irradiation of the light-shift beam,
we measured $\sigma_{+}$ and  $\sigma_{-}$ counts for the cases with and without irradiation.
Figure 3 shows the scatter plots of $\sigma_{+}$ and  $\sigma_{-}$ counts for the $\ket{\uparrow}$ (red circles) and $\ket{\downarrow}$ (blue squares).
The average number of photon counts for a single atom transit is approximately 5.5 counts/atom.
In this measurement, a few atoms pass through the cavity mode during a single measurement window of 2 ms,
so that the number of photon counts obtained is sufficiently high to neglect the influence of dark counts at the two detectors.
It should be noted that the mean intracavity atom number is much less than unity.
The frequency of the excitation beam is set to each resonance value of the cavity-enhanced spectra obtained in Fig. 2(b) and (c).   
When the light-shift beam is not injected into the cavity,
$\sigma_{+}$ and $\sigma_{-}$ counts are positively correlated with the correlation coefficient of 0.58 and 0.70 for $\ket{\uparrow}$ and $\ket{\downarrow}$, respectively [Fig. 3 (a)].
The same number of $\sigma_{+}$ and $\sigma_{-}$ photons are emitted regardless of the spin states of atoms.
This is because of the random spin flip induced by the degenerated all substates of $^{3}P_{1} (F' = 3/2)$.
When the light-shift beam is injected into the cavity [Fig. 3 (b)],
the correlation between $\sigma_{+}$ and $\sigma_{-}$ counts is dramatically reduced, 
where the correlation coefficients are 0.38 and 0.11 for $\ket{\uparrow}$ and $\ket{\downarrow}$, respectively; that is,
the polarization of emitted photons strongly depends on the initial spin state.

\begin{figure}[t]
   \includegraphics[width=7.5cm]{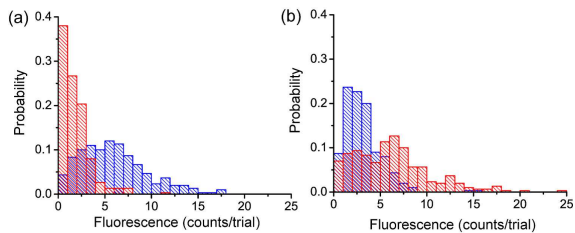}
\caption{Projective measurements of nuclear spin qubit.
The initial states are (a) $\ket{\uparrow}$ and (b) $\ket{\downarrow}$.
The blue and red bars indicate $\sigma_{+}$ and  $\sigma_{-}$ counts, respectively.
}
\label{fig4}
\end{figure}

\begin{figure}[t]
\includegraphics[width=7.5cm]{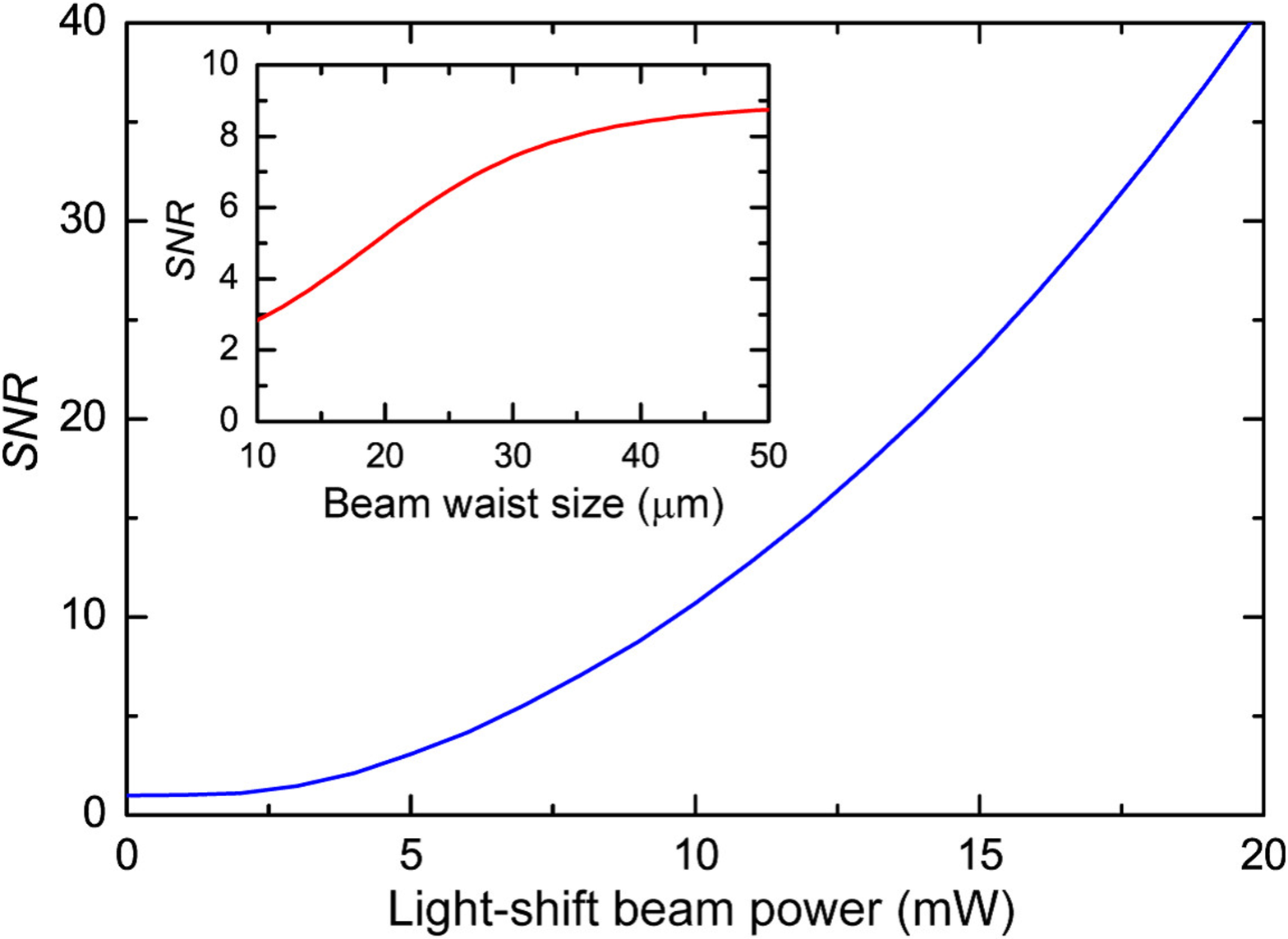}
\caption{(color online) Theoretically calculated $SNR$.
A Gaussian profile with $w_{l}=50$ $\mu m$ is assumed.
The inset shows the calculated $SNR$ versus $w_{l}$, where the peak intensity of the light-shift beam is constant for all $w_{l}$.}
\label{fig5}
\end{figure}

Figure 4 shows the results of projective measurements at the level of a single atom.
The spin states are $\ket{\uparrow}$ and $\ket{\downarrow}$ for Fig. 4(a) and (b), respectively.
The average transit atom number is approximately 1.0 and 1.2 per single measurement for the $\ket{\uparrow}$ and $\ket{\downarrow}$ states.
We define the signal to noise ratio ($SNR$) of the projective measurement as the desired counts divided by the undesired counts, e.g., $\sigma_{+}/\sigma_{-}$ counts for the initial spin $\ket{\uparrow}$.
The SNRs for the initial spin $\ket{\uparrow}$ and $\ket{\downarrow}$ are 4.2 and 2.2, respectively.

We simulated the $SNR$ using the master equation with adiabatic approximation \cite{Peng10}.
In this simulation, the influence of dark counts at the detectors is not taken into account.
Figure 5 shows the theoretically predicted $SNR$ as a function of the light-shift beam power.
Although a further improvement in the $SNR$ is expected by increasing the available intensity of the light-shift beam,
there is a discrepancy between the predicted $SNR$ of 8.7 and the experimental values of 5.1 and 3.5 obtained after the subtraction of the dark counts at the detectors. 
The inset of Fig. 5 shows a graph of the predicted $SNR$ versus $w_{l}$.
From this graph, one can see that the $SNR$ is sensitive to variation in the spatial intensity.
Thus, we can qualitatively explain the discrepancy between the theoretical and experimental values by the additional intensity modulation and the misalignment of the propagation axes of the counter-propagating beams. 


In conclusion, we have reported the construction of a two-mode cavity QED system,
in which the ac Stark effect of the $^{3}P_{1} (F' = 3/2)$ state and $^{3}D_{1} (F' = 1/2)$ states is used to avoid the unwanted spin flips.
The polarization of emitted photons depends on the nuclear spin state.
By utilizing fluorescence having a nuclear-spin-dependent polarization,  
we have performed a projective measurement on a single nuclear spin of $^{171}$Yb atoms.
Our method, described in this paper, will facilitate applications in quantum information communication technology and metrology, such as quantum networks using the nuclear spin and spectroscopic measurements for the alkaline-earth-like atoms.

We would like to thank M. Takeuchi, N. Takei, T. Mukaiyama, T. Kishimoto, S. Inouye, K. Hayasaka, and A. Yamaguchi for their stimulating and fruitful discussions.
We would also like to thank H. Kanamori, Y. Awaji, and T. Hirano for supporting us during the experiments.


\begin{thebibliography}{99}

\bibitem{Ye08}
J. Ye, H. J. Kimble, and H. Katori,
Science \textbf{320}, 1734 (2008). 

\bibitem{Reichenbach07}
I. Reichenbach and I. H. Deutsch,
\prl \textbf{99}, 123001 (2007).

\bibitem{Reichenbach09}
I. Reichenbach, P. S. Julienne, and I. H. Deutsch,
\pra \textbf{80}, 020701(R) (2009).

\bibitem{Daley08}
A. J. Daley, M. M. Boyd, J. Ye, and P. Zoller,
\prl \textbf{101}, 170504 (2008).

\bibitem{Gorshkov09}
A.V. Gorshkov, \textit{et al.},
\prl \textbf{102}, 110503 (2009).

\bibitem{Shibata09}
K. Shibata, \textit{et al.},
Appl. Phys. B \textbf{97}, 753 (2009).

\bibitem{Gorshkov10}
A.V. Gorshkov, \textit{et al.},
Nature Physics \textbf{6}, 289 (2010).

\bibitem{Fukuhara07}
T. Fukuhara, S. Sugawa,M. Sugimoto,S. Taie, Y. Takahashi,
\pra \textbf{79}, 041604(R) (2009).

\bibitem{Kimble98}
H. J. Kimble,
Physica Scripta \textbf{T76}, 127 (1998).

\bibitem{Vahala03}
K. J. Vahala,
Nature \textbf{424}, 839 (2003).

\bibitem{Takeuchi10}
M. Takeuchi, \textit{et al.},
\pra \textbf{81}, 062308 (2010).

\bibitem{Noguchi10}
A. Noguchi, Y. Eto, M. Ueda, M. Kozuma,
arXiv:1005.3584.

\bibitem{Takei10}
N. Takei, \textit{et al.},
\pra \textbf{81}, 042331 (2010).

\bibitem{Hu09}
C. Y. Hu, W. J. Munro, J. L. O'Brien, and J. G. Rarity,
\prb \textbf{80}, 205326 (2009).

\bibitem{Dehmelt82}
H. Dehmelt,
IEEE Trans. Instrum. Meas. \textbf{31}, 83 (1982).

\bibitem{Wang07}
Y. H. Wang, \textit{et al.},
Laser Phys. \textbf{17}, 1017 (2007).

\bibitem{Kohno09}
T. Kohno, \textit{et al.},
Appl. Phys. Express \textbf{2}, 072501 (2009). 

\bibitem{Lemke09}
N. D. Lemke, \textit{et al.},
\prl \textbf{103}, 063001 (2009).

\bibitem{DeMile95}
D. DeMille,
\prl \textbf{74}, 4165 (1995). 

\bibitem{Loftus01}
T. H. Loftus, Ph.D. thesis, University of Oregon, 2001.

\bibitem{Peng10}
The evolution of the atomic density matrix $\rho \left(t\right)$ is assumed to be $\rho \left( t\right) =p_{a}\left(t\right) \rho_{sa}
+p_{b}\left(t\right) \rho _{sb}$, where $\rho _{sa}$ is the normalized projected steady state of the Liouville operator in the subspace spanned by the state $|\downarrow \rangle $, the $^{3}P_{1}$ states with $m_{F'} = -3/2, 1/2$, and the $^{3}D_{1}$ state with
$m_{F'} = 1/2$, and $\rho _{sb}$ is that for the subspace spanned by the four other atomic states. 
Under such an approximation we derive the rate equation of the probabilities $p_{a,b}\left(t\right) $ of the atom in the two subspaces and then obtain the atomic density matrix $\rho \left( t\right)$.

\end{thebibliography}
\end{document}